\documentclass[12pt,a4paper]{article}
\usepackage{epsf,amsmath}
\usepackage{makeidx}  
\usepackage{graphicx} 

\newcommand{\rf}[1]{(\ref{(1)})}

\def \be {\begin{equation}}
\def \ee {\end{equation}}
\def \HM {HEIDEL\-BERG-MOSCOW~}

\def \00 {$0^+\to 0^+$}

\newcommand{\ba}[1]{\begin{eqnarray} \label{(1)}}
\newcommand{\ea}{\end{eqnarray}}

\begin{document} 
\title{WHY IS THE CONCLUSION OF THE GERDA EXPERIMENT NOT JUSTIFIED ?}

\author{{\it H.V. Klapdor-Kleingrothaus $\footnote{prof.klapdor-kleingrothaus@hotmail.de}$ 
and I.V. Krivosheina $\footnote{irinakv57@mail.ru}$ }\\ 
				Heidelberg,  Germany
}

\maketitle

\begin{abstract}
   The first results of the GERDA  double beta experiment in Gran Sasso 
   were recently presented. 
   They are fully consistent with the \HM experiment, 
   but {\it because of its low statistics cannot proof anything at this moment}.
   It is no surprise that the statistics is still far from being able 
   to test the signal claimed by the \HM experiment. 
   The energy resolution of the coaxial detectors is a factor of 1.5 worse 
   than in the \HM experiment. 
   The {\it original goal} of background reduction 
   to $10^{-2}$\,counts/kg\,y\,keV, or by an order of magnitude 
   compared to the \HM experiment, {\it has not been reached}. 
   The background is {\bf only} a factor 2.3 lower if we refer 
    it to the experimental line width, i.e. in units 
    counts/kg\,y\,energy resolution. \\
 
     {\it With} pulse shape analysis (PSA) the background 
     in the HEIDELBERG-MOSCOW experiment 
     around Q$_{\beta\beta}$ is 4$\times$10$^{-3}$counts/kg\,y\,keV 
\cite{HVKK-IVK-MPhLA2006}, 
			which is {\it a factor of 4 (5 referring to the line width) lower} 
     than that of GERDA with pulse shape analysis.\\
 
     The amount of enriched material used in the GERDA 
     measurement is 14.6\,kg,  
     only a factor of 1.34 larger than that used in the \HM experiment.  
     The background model is oversimplified and not yet adequate.  
     It is not shown that the lines of their background can be identified.  
     GERDA has to continue the measurement further $\sim$5\,years, 
     until they can responsibly present an understood background.
     The present half life limit presented by GERDA of 
     T$_{1/2}^{0\nu}>$2.1$\times$10$^{25}$\,y (90\% confidence 
     level, i.e. $1.6\sigma$) is still lower than the half-life of 
     T$_{1/2}^{0\nu}=$2.23$^{+0.44}_{-0.31}${$\times$ 10$^{25}$}\,y 
\cite{HVKK-IVK-MPhLA2006} 
     determined in the \HM experiment.  \\   



\end{abstract}

\section{INTRODUCTION}

   Nuclear double beta ($\beta\beta$) decay is one of the flagships 
   of non-accelerator particle physics searching for beyond standard model physics underground 
\cite{KK70Y}. 
   For many years (since 1992) the \HM experiment 
   using the first enriched high-purity $^{76}${Ge} detectors dominates 
   the field in sensitivity 
\cite{CERN12-CERN02}. 

      However, recently some fresh breeze  arose in the field. 
      The EXO and the Kamland-Zen experiments looking 
      for $\beta\beta$ decay of $^{136}${Xe} 
      reached half-life limits of order of $10^{25}$\,years 
      (1.6 and $1.9 \times 10^{25}$\,years (90\% c.l.), respectively) 
\cite{EXO-Kamland12}. 
     These results are consistent with \HM  
\cite{HVKK-IVK-MPhLA2006} 
     within 1 or 2$\sigma$ with the matrix elements of 
\cite{St-Mut-KK90}. 
      They unfortunately suffer, however from low energy resolution 
      ($\sim$30 times less than Ge detectors). 

    These days the GERDA experiment in Gran Sasso reported its first results 
\cite{LNGS-SEmin2013,Catt2013}.  
    It used the idea of the GENIUS Project 
\cite{HVKK-JH-MH97-ProjGen} 
     namely installing naked Ge detectors in liquid nitrogen or liquid argon. 
     Operating 14.6\,kg of enriched $^{76}${Ge} --- of them 10\,kg from 
     the \HM experiment --- GERDA derived 
     after exposure of 1.5\,y a lower limit of 
     T$_{1/2}^{0\nu} > 2.1\times 10^{25}$\,y 
     at 90\% confidence limit (1.6 $\sigma$) 
     from their pulse-shape selected spectrum. 
     On this basis they claim 'refuting of the \HM 
     signal at high probability'    
\cite{LNGS-SEmin2013}.
     To this experiment and this conclusion we have 
     the following comments.

\section{GENERAL DEFICIENCIES OF THE \\EXPERIMENT AND THE ANALYSIS}
\subsection{HALF LIFE AND DOUBLE LINE}

     The conclusion that the result of the \HM experiment 
     (a signal at a 6.4 $\sigma$ c.l.  
\cite{HVKK-IVK-MPhLA2006}) 
      is refuted, is {\bf wrong}. 

{\it The reason is very simple.} 
     The authors of  
\cite{LNGS-SEmin2013,Catt2013}, 
      compare their result to the line at 
      (2038.1-2038.5)$\pm0.5$(stat.)$\pm1.2$(syst.)\,keV 
     in the {\bf full} (not PSA - treated) spectrum taken 
     by \HM (exposure 71.7\,kg\,y) see
\cite{KK04-PL}.
     They ignore the result given in 
\cite{HVKK-IVK-MPhLA2006}, 
      that this  line at Q$_{\beta\beta}$ in the \HM experiment 
      is {\it a  double}  line which cannot be resolved 
      by the energy resolution of a Ge detector. 
      {\it Two lines} of almost equal intensity occur at 2037.5 
      and 2039.3\,keV. 
      They can be separated, however, by pulse shape analysis (PSA), 
      since the first one consists essentially of single site 
      events as expected for a 0$\nu\beta\beta$ line, 
      the second essentially of multiple site events, as expected 
      for a gamma line.
      Their intensities are 11.0$\pm$1.8 and 10.3$\pm$3.3, 
      respectively, adding to the line found in the full 
      (not PSA-treated) spectrum of 19.6$\pm$5.4 events 
      obtained after 51.39\,kg\,y (the time during which the time 
      structure of pulses has been recorded in the \HM experiment).  
      The 0$\nu\beta\beta$ half life is consequently 
      $(2.23^{+0.44}_{-0.31}) \times 10^{25}$\,years 
\cite{HVKK-IVK-MPhLA2006}, 
      and not (${1.19}^{+0.38}_{-0.22})\times 10^{25}$\,y 
      as deduced in 
\cite{KK04-PL}               
      from the full spectrum 
      and as assumed in the GERDA report. 

      Therefore, in their window around Q$_{\beta\beta}$ GERDA 
       should expect 3.1$\pm0.8$\,events
      only (but {\it not} $5.9\pm 1.4$ ($1\sigma$ error!) as they claim). 
      This signal should be searched at an energy 
      of 2037.5\,keV$\pm$0.5(stat.)$\pm$1.2(syst.), where 
      the single site line is found, {\it not} at 2039\,keV where 
      we observe the multiple $\gamma$-line.

      The fact that the line at Q$_{\beta\beta}$ in the full 
      spectrum in the \HM experiment is a double line, has been observed       
      independently also by I. Kirpichnikov 
\cite{Kirp2010}.
       He found further indication of a tiny third line at $\sim$2034.5\,keV, 
       unresolved from the line seen at 2038.5\,keV.
       It shows up also in our spectrum of multiple site events rejected 
      by PSA (see Fig.5 and Fig.6 of DARK2007 Proc. 
\cite{HVKK-IVK-MPhLA2006,KK70Y}).
       He claimed that the lines at 2039.3\,keV and 2034.5\,keV 
       each are a sum line of two consecutive gamma transitions 
       (as it is the case also for the 2016.7\,keV line -- see below and
\cite{NIM03-HVKK}). 
       He also showed that this is supported by the GEMMA Experiment 
\cite{Beda2004}. 
      So their existence does not contradict to the findings of Gromov et. al 
\cite{Grom2006} 
      and D\"orr et al. 
\cite{Doer-HVKK03} 
       that there is {\bf no} gamma transition of this energy 
       in known radioactive isotopes.                

	Concluding GERDA has to compare its $1.6\sigma$ limit 
	of $2.1\times10^{25}$\,y to the \HM $6.4\sigma$ signal 
	of single site events yielding 
	T$_{1/2}=(2.23^{+0.44}_{-0.31})\times10^{25}$\,y
\cite{HVKK-IVK-MPhLA2006}. 
        This means that the GERDA limit is lower than the \HM half life, 
        and is fully consistent with \HM, 
        but {\it because of its low statistics GERDA cannot proof anything 
        at this moment}.

\begin{figure}[!b]
\begin{center}
\includegraphics[scale=1.1]{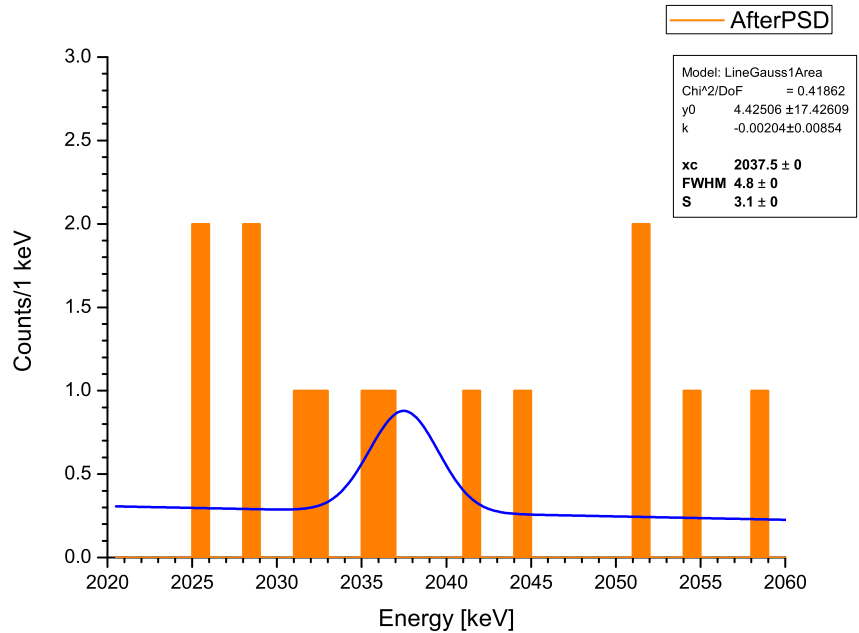}
\caption[*]{Gerda spectrum {\it after} pulse shape discrimination (PSD), from 
\cite{LNGS-SEmin2013,Catt2013}.
 						The solid line corresponds to the correct expectation from the 
 						\HM experiment 
\cite{HVKK-IVK-MPhLA2006,KK70Y}
 						(see text)}
\end{center}
\label{fig1}
\end{figure}
 

       In their slides 54-57 
\cite{LNGS-SEmin2013}     
       they show their 90\% upper limit. This is {\it not} 
       a fit, but simple superposition of a Gaussian line with fixed energy, 
       intensity and width on the background.
       In the upper parts of slides 56, 57 
\cite{LNGS-SEmin2013}
       they compare this to their expectations from the erroneously assumed half life  
        for the \HM experiment 
\cite{KK04-PL} 
        -- and {\it not} at the energy position where the line has been observed. 
       The {\it correct} expectation from \HM  
\cite{HVKK-IVK-MPhLA2006} 
        is shown in Fig.1 of this paper, which should replace the dotted red line 
        in the slides 56, 57 of 
\cite{LNGS-SEmin2013}.          
        In Fig.1 from the 5 parameters of the fit program all, {\it except the two
        background parameters}, are fixed: position of the line and its expected 
        intensity {\it according to}  
 \cite{HVKK-IVK-MPhLA2006},
        FWHM according to GERDA experimental resolution.


    {\sf No contradiction to the GERDA result can be seen here.}

            But even the discrepancy  between the limit of 
	          T$_{1/2}>2.1\times10^{25}$\,y (90\% c.l., i.e. $1.6\sigma$ level) 
						of GERDA and the half-life of $1.19\times10^{25}$\,y, 
						which they assumed erroneously,  
            is less than the $2\sigma$ uncertainty 
				of this half life (which is ($1.19^{+1.08}_{-0.39}\times10^{25}$\,y).
				Already this certainly would {\bf not} justify the strong statement 
				of refuting the \HM result.

\subsection{BACKGROUND AND\\ BACKGROUND MODEL BUILDING}

     It has been mentioned that {\it the main goal} of reaching a background 
     of 10$^{-2}$counts/kg\,y\,keV, or by an order of magnitude lower 
     than in the \HM experiment, was {\it not reached} by GERDA. 
     The background of GERDA in the energy window 2000-2060\,keV 
     around  Q$_{\beta\beta}$ is 0.031\,counts/kg\,y\,keV. 
     This is a factor of 3.5 lower than the \HM background 
     of $0.113\pm0.007$\,counts/kg\,y\,keV 
\cite{KK04-PL}. 
     It is, however, only by a factor of 2.3 smaller than \HM if we refer 
     to the line width, i.e. in units of counts/kg\,y energy resolution. 
     It is the latter value which defines the sensitivity of the experiment 
     concerning background (without PSA). 
     {\it With} pulse shape analysis the background in the \HM experiment 
     is $\sim$$4\times10^{-3}$\,counts/kg\,y\,keV
\cite{HVKK-IVK-MPhLA2006}  
     around Q$_{\beta\beta}$ (range 2000-2060\,keV). This is 
     {\it a factor of 4 (5, when referring to the line width) lower} 
     than found by GERDA with PSA.

    A general criticism has to be made concerning the treatment 
    of background by GERDA, which is completely unsufficient at this moment.
    The main point is that they do not show that all lines 
    in the spectrum are understood.
    The spectra are shown in 
\cite{LNGS-SEmin2013,Catt2013}
     -- except for a range of $\sim$ 1900--2200\,keV around 
    Q$_{\beta\beta}$ -- {\it only binned into energy bins of 5\,keV}. 
    This is hardly adequate to an energy resolution of 4.8\,keV. 
    More critical, is that they compare most part of their spectra 
    {\it in a 30\,keV binning} to a background model averaging also over 30\,keV.
     This means that for most part of the spectrum individual gamma 
     lines are not shown and their intensities were not determined 
     or at least are not listed. 
     In this way the usual procedure of localizing the sources 
     of radioactive impurities in the setup cannot be applied. 
     Further, in such way it cannot be checked whether there exist 
     lines in the spectrum not included in their background model. 
     Strange is also, that they (their slide 55, from   
\cite{LNGS-SEmin2013}
      ) exclude 'lines' at 2104 and 2119\,keV from 
     their background - at least the first of them being not visible. 
     If the 2104 and 2119\,keV lines are accepted as lines than 
     there are {\it m a n y}  lines in their 2\,keV-binned spectrum to be accepted 
     and to be explained.  
                                               
     Also about the background in the {\it i n d i v i d u a l}  detectors 
     nothing can be said in this way.       
     It is clear that under such circumstances the background 
     cannot yet be claimed to be understood. 

     In the \HM experiment more than 70 lines have 
     been seen and {\it identified} in the spectrum 
\cite{Doer-HVKK03,Grom2006}. 
      A special investigation of the background around Q$_{\beta\beta}$, 
      and of the intensity ratios of the $^{214}$Bi lines, 
      some of which occur in the window around Q$_{\beta\beta}$ 
      (range 2000 - 2100\,keV), has been performed with a $^{226}$Ra source 	
\cite{NIM03-HVKK}. 
       In particular the effect of true coincidence summing on the intensities 
       has been studied, in particular, for the line at 2016.7\,keV, which 
       as E0 transition can be seen only as sum line of two consecutive gamma transitions -- 
       as the line identified at 2039.3\,keV  (see above). 

       {\sf Concerning their background models:}
       The comparison between measured  spectrum  and calculated 
       background in the GERDA reports, in spite of the smoothening 
       of the data in 30\,keV bins -- which {\it suppresses} local deviations -- 
       shows differences up to a factor of 2 and more (a factor of 2.3 in the region 
       around Q$_{\beta\beta}$ !) (Slides 33, 34 in 
\cite{LNGS-SEmin2013}). 
       This raises the question whether these models are sufficient.

				Highly surprising is the statement that a socalled minimal background model  
				${\bf not}$  including lines from  $^{214}$Bi should be sufficient. 
				$^{214}$Bi was found in the \HM experiment to yield the  famous  
				lines closest to Q$_{\beta\beta}$ 
\cite{KK04-PL,NIM03-HVKK}, 
        and is clearly seen also by GERDA 
				(e.g. at 2204\,keV), and seems to show up also already in the region around 
				Q$_{\beta\beta}$. 

\begin{figure}[!htb]
\begin{center}
\includegraphics[scale=1.1]{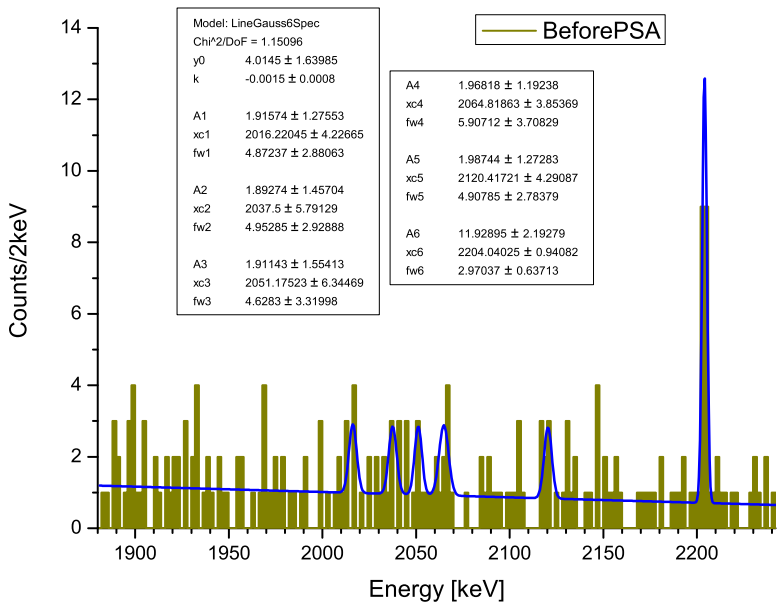}
\caption[*]{Gerda spectrum {\it before} PSD with 2\,keV - binning (from 
\cite{LNGS-SEmin2013,Catt2013}). 
						Besides the strong $^{214}$Bi line at 2204\,keV,  
						our fit finds indications of lines (on 1-2$\sigma$ level) 
						at known positions of $^{214}$Bi lines 2016.7\,keV, 
						(not separable from 2010.7 and 2021.8\,keV), 2052.9\,keV,
						2119\,keV. Further it finds a line at 2037.5\,keV 
						(not separable from lines 2034.5 and 2039.3\,keV). 
						(The parameter 'A' of the fit is connected with the intensity 'N' 
						by N=S/2, S=A$\times$(FWHM/2)$\times\sqrt{\pi/{ln2}}$. 
						The other parameters are self-explainable (see text))
						}
\end{center}
\label{fig2}
\end{figure}

        Fig. 2 shows a fit which could complement their full spectra shown 
        in slides 54, 55, 56 (lower part) of 
\cite{LNGS-SEmin2013}. 
 
				Besides the strong $^{214}$Bi line at 2204\,keV, some lines of low statistics 
				(between 1 and 2$\sigma$) are indicated in the range around Q$_{\beta\beta}$ 
				(2000-2120\,keV) at energies corresponding to $^{214}$Bi lines (2016.7\,keV, 
				not resolvable from 2010.7 and 2021.9\,keV Bi lines), 2052.9\,keV, 2119\,keV. 
				A line at 2065\,keV is not understood. Also a line at 2037.5\,keV is indicated, 
				not separable from 2034.5 and 2039.3\,keV lines. Its intensity of 4.9$\pm$3.8\,counts 
				is consistent with the expectation from \HM 
\cite{HVKK-IVK-MPhLA2006,KK70Y},
				which is (4.4$\pm$1.0) counts, adding 
				the expectations from the 2037.5\,keV line and for half of the 2039.3\,keV line 
				(because of lower background of GERDA without PSD). 
				In case GERDA would not see the 2039.3 keV\,line, the expected value from \HM 
				would be 3.1$\pm$0.8, both being fully consistent with the above given fit. 
				Considering the {\it number of events} to be expected in the GERDA spectrum before PSD, 
				in a region of 2 FWHM (4.6$\sigma$) around 2037.5\,keV, a value 
				of 8.9$\pm$1.8\,events is expexted,with the background of 0.5 counts per keV 
				determined by the fit in Fig.2. 
				The observed value of events in this range is 10. 
				So the present GERDA result is fully consistent with the result of \HM 
\cite{HVKK-IVK-MPhLA2006,KK70Y}.


\begin{figure}[!ht]
\begin{center}
\includegraphics[scale=1.1]{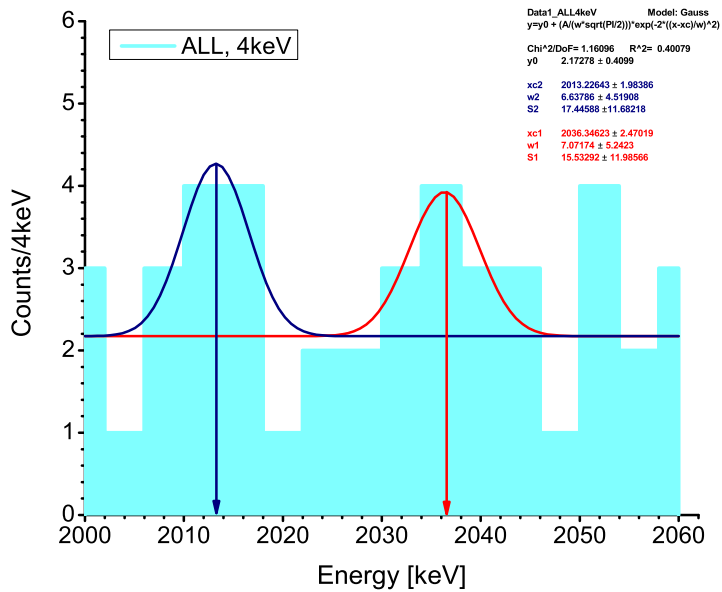}
\caption[*]{The Gerda spectrum {\it before} PSD, with 4\,keV - binning 
						and our fit. Two structures arise around 2014\,keV
						(unresolved $^{212}$Bi - 2010.7, 2016.7, 2021.8\,keV), 
						and a broad unresolved structure covering the range 2034-2040\,keV
						(lines 2034.5, 2037.5, 2039.3\,keV). 
						(In the fit 'S' determines the number of counts N in the line, N=S/4)}
\end{center}
\label{fig3}
\end{figure}

        Fig. 3 shows the full GERDA spectrum (before PSA) with a binning of 4\,keV. 
        Two structures dominate the spectrum, a broad line at the location 
        of the unresolved $^{214}$Bi lines at 2010.7, 2016.7 and 2021.8\,keV 
        (known from the \HM experiment) 
\cite{KK04-PL}, 
        and a broad unresolved structure covering the range 2034 to 2040\,keV, 
        which includes the lines 2034.5, 2037.5 and 2039.3\,keV known from \HM 
\cite{HVKK-IVK-MPhLA2006}.
        The fit yields for the second line E=2036.4$\pm$2.5\,keV and an intensity 
        of 3.9$\pm$3.0\,events, consistent with the expectation (see above) 
        of 4.4$\pm$1.0\,events. {\sf A word of caution:} It should be mentioned,
        however, that different ways of 4\,keV binning can give rather
        different result. Thus Figs. 2 and 3 show, that at the low statistics 
        GERDA has at present, and with the low energy resolution 
        the GERDA detectors have, at this moment it is premature and marginal 
        to search for resolved lines in this region. 
        The {\it statistics} of GERDA at present is {\it simply not sufficient} 
        to check the result of the \HM experiment.
        But what is found, 
        is consistent with the expectation from \HM 
\cite{HVKK-IVK-MPhLA2006,KK70Y}.

        The GERDA report claims no excess of signal counts above background.
        However, even according to their Table (Slide 52 in 
\cite{LNGS-SEmin2013}) 
        an excess {\it is there}.

					Unusual is that the authors show and determine the background spectrum 
					not over the full measuring time (November 2011 until May 2013) 
					but only until January 2013 
\cite{LNGS-SEmin2013,Catt2013,Knopf2013}. 
          The reason should be given: Why 4-5\, months of statistics remained unused?
          One would need further some proof that the background during the period 
          January to May 2013 was the same as in the period before!

\subsection{DETECTOR RESOLUTION AND \\ STABILITY OF ELECTRONICS}
					The energy resolution and its time stability of the coaxial detectors 
              through 1.5\,years of operation is rather modest, the energy 
              of the 2614.5\,keV Th line floating between -1.4 and +2.5\,keV 
							around the average value. 
					The average resolution lies for the different detectors 
              between 4.2 and 5.8\,keV, and averaged over the detectors is 4.8\,keV. 
					This is a factor of 1.5 worse than  the resolution of the {\it same} 
							detectors during eight years of measurement 
							in the \HM  experiment, which was 3.27\,keV 
\cite{KK04-PL}. 
              The reason should be given.
					This could also indicate some time instability of the electronics 
              in the GERDA experiment. 
					No analysis of a possible temperature dependence of the setup 
							and electronics is mentioned.                             

\subsection{PULSE SHAPE ANALYSIS}
 
          The training of the neuronal net for GERDA is done with 
          the method given in 
\cite{KK99-PS}
          using the 1592.5\,keV double escape line of the 2614.5\,keV $^{228}$Th
     			line for simulating single site events, and the 1612\,keV total
     			absorption peak from the $^{228}$Th daughter nuclide $^{212}$Bi 
     			for multiple site events. 
				Unfortunately the time structure of all individual events 
				in the relevant range of energy around Q$_{\beta\beta}$ is {\it not} 
				shown by GERDA. 
				The reduction of the $\gamma$-background is rather modest (order of factor 2).

					There are, however, still differences between 0$\nu\beta\beta$ 
					(and 2$\nu\beta\beta$) events and DE--$\gamma$--events 
					in time structure and size (partial volumes in the Ge detector 
					inside which the energy of the events is released),  
					see Monte Carlo simulations in 
\cite{KK-IT-IV}.

					These differences of 0$\nu\beta\beta$ events of different effective 
					neutrino mass $m$ and right-handed current parameters $\theta, \lambda$ 
					from single site (DE) $\gamma$ events are such, that even with 
					the typical spatial resolution of a large Ge detector it might 
					not be excluded to separate 0$\nu\beta\beta$ events sharper 
					from any kind of $\gamma$-event, if the neuronal net 
					could be properly 'calibrated'.

					In 
\cite{HVKK-IVK-MPhLA2006}
					this has been tried in some empirical way, with the result of a drastic 
					further reduction of the {\it whole} $\gamma$-background 
					to $\sim$4$\times$10$^{-3}$\,counts/kg\,y\,keV (see 
\cite{HVKK-IVK-MPhLA2006} 
					and also Fig.33 (left) in 
\cite{KK04-PL}). 
          With this reduction the candidate 0$\nu\beta\beta$ line stands
          out clearly of the background.

\subsection{PROBLEMS WITH DETECTORS IN LIQUID ARGON?}
 
	Despite GERDA operated its detectors in liquid argon in shrouds of very 
	thin copper (does this mean that it is tried to {\it not} use naked detectors?) 
	already two detectors could not be used. 
	This led to the fact that instead of 17.7\,kg of enriched material 
	only 14.6\,kg have been used (not much 
	more than  the 10.9\,kg used in the \HM experiment). 
	As reason was given too high  leakage current. 
	Nothing is said in the report about the behaviour 
	of the leakage currents of the other detectors as function of running 
	time in the liquid argon. 

	The experience from our GENIUS Test Facility, in which 
	we operated six (non-enriched) naked Ge detectors over 
	the period of three years in Gran Sasso was  the following 
\cite{HVKK2003-04-06NIM,KK06-08}: 
         Limited long-term stability of naked detectors in liquid nitrogen 
	 as a result of increasing leakage current. After three years none 
	 of the six detectors was working any more with the nominal 
	 leakage current. 
	 Three of the detectors did not work any more at all.   

\section{CONCLUSION}

            The most sensitive double beta decay experiments {\it at present 
						under operation}, EXO and KAMLAND-ZEN 
\cite{EXO-Kamland12}
              and GERDA 
\cite{LNGS-SEmin2013,Catt2013}
              reported lower limits for 
              neutrinoless double beta decay on a 90\% c.l., which are consistent  
              with the 6.4$\sigma$ \,signal delivered 
							by the \HM Experiment 
\cite{HVKK-IVK-MPhLA2006}. 
              All of these experiments plan improvements of their sensitivity. 
							The future SNO+ experiment with the $\beta\beta$ emitter $^{130}$Te 
							still is only under discussion 
\cite{SNO-2013}.   
              It is obvious, that it will take quite some more years, until checking 
              of the \HM positive result will become possible. 
              
              In the case of EXO and KAMLAND-ZEN the modest energy resolution of
              $\sim$ 90\,keV may make serious problems in the moment where indications 
							of a signal might be found - remember the lines close to 
              the 0$\nu\beta\beta$ line from \HM. 

              {\sf In the case of GERDA:}
 \begin{enumerate}             
              \item  should be improved considerably the treatment of 
                   background data, which at present is on an unacceptable level. 
                   It is not acceptable that no list of identified lines and intensities exists 
                   and consequently no comparison with expectations. Consequently 
                   no satisfactory localisation of the radioactive impurities 
                   in the setup could be performed. 
                   It is further not acceptable that lines remain
                   unexplained in the spectrum.
              \item  the statistics of the experiment has to be decisively improved before 
                   any relevant statements can be made, to avoid premature conclusions as 
                   in the present report.
              \item  the reasons for the limited energy resolution of the detectors 
                   have to be explained, and the resolution has to be improved 
                   to an acceptable level.
              \item  the time structure of their events in the relevant energy 
										range around Q$_{\beta\beta}$ 
										should be individually shown, and also their fits 
                   	by their pulse shape approximation library (as has been done in 
\cite{HVKK-IVK-MPhLA2006}).
              \item  In view of the experience with GENIUS-TF 
\cite{HVKK2003-04-06NIM,KK06-08} 
										the development of the leakage currents of the detectors 
										as function of time should be shown.
                   After two detectors already did not work because of too high 
										leakage current in the present run of GERDA, it should be made 
										sure that not more detectors will be lost by the operation in liquid argon.
\end{enumerate}

             Because of its {\sf similar background and detector mass} 
             {\it GERDA would require similar measuring times} as \HM 
             to get comparable statistics.   
 
					Some outlook on the future of $\beta\beta$ experiments is given in 
\cite{KK06-08}.

\section{ACKNOWLEDGEMENTS}

       The authors gratefully acknowledge the important contribution 
        of Dr. S.N. Karpov to this paper.
				It is our pleasure to give our deeply felt thanks here to all friends and
				colleagues, who have supported us so efficiently in various ways on our way 
				through double beta decay research during the last twenty-five years.



\end{document}